\documentclass[aps,prl,twocolumn,superscriptaddress,10pt]{revtex4-2}
\usepackage[utf8]{inputenc}
\usepackage{mathrsfs}
\usepackage{dsfont}
\usepackage{amsmath}
\usepackage{amssymb}
\usepackage{amsthm}
\usepackage{amsfonts}
\usepackage{amstext}
\usepackage{amsopn}
\usepackage{amsxtra}
\usepackage{dsfont}
\usepackage{color}
\usepackage{graphicx}
\usepackage{hyperref}
\usepackage{xurl}
\usepackage{tikz}
\usepackage{enumerate}
\usepackage[capitalize]{cleveref}
\usepackage{datetime2}

\usepackage{empheq}

\usepackage{esint}

\newtheorem{theorem}{Theorem}

\newcommand{\R}{\mathbb{R}}
\newcommand{\Z}{\mathbb{Z}}

\newcommand{\ii}{\infty}

\renewcommand\phi{\varphi}

\newcommand{\cP}{\mathcal{P}}

\newcommand{\cV}{\mathcal{V}}

\newcommand{\cB}{\mathcal{B}}

\newcommand{\cH}{\mathcal{H}}

\newcommand\pscal[1]{{\ensuremath{\left\langle #1 \right\rangle}}}
\newcommand{\norm}[1]{ \left\| #1 \right\|}
\newcommand{\nnorm}[1]{ \left\| #1 \right\|'}

\renewcommand{\geq}{\geqslant}
\renewcommand{\leq}{\leqslant}

\renewcommand{\tilde}{\widetilde}

\newcommand{\nn}{\nonumber}
\newcommand{\rd}{\mathrm{d}}

\newcommand{\dt}{\rd t}
\newcommand{\br}{\mathbf{r}}
\newcommand{\bR}{\mathbf{R}}
\newcommand{\bk}{\mathbf{k}}
\newcommand{\bK}{\mathbf{K}}

\let\Re\relax
\let\Im\relax
\DeclareMathOperator{\Re}{Re}
\DeclareMathOperator{\Im}{Im}

\begin{document}

\title{Time-Dependent Density Functional Theory with Coulomb Interactions}


\author{Asbj{\o}rn~B{\ae}kgaard~Lauritsen}
\email{lauritsen@ceremade.dauphine.fr}
\affiliation{CEREMADE, CNRS, Université Paris-Dauphine, PSL Research University, Place de Lattre de Tassigny, 75016 Paris, France}

\author{Mathieu~Lewin}
\email{Mathieu.Lewin@math.cnrs.fr}
\affiliation{CEREMADE, CNRS, Université Paris-Dauphine, PSL Research University, Place de Lattre de Tassigny, 75016 Paris, France}

\author{Jakob~Oldenburg}
\email{jakob.oldenburg@cnrs.fr}
\affiliation{CEREMADE, CNRS, Université Paris-Dauphine, PSL Research University, Place de Lattre de Tassigny, 75016 Paris, France}

\begin{abstract}
We provide the first fully rigorous justification of time-dependent density functional theory in the continuum: 
Given a time-dependent density, we prove that there exists an external potential, unique up to a time-dependent constant, such that the corresponding Schrödinger equation reproduces the given density. This potential can be obtained with an iteration scheme. 
Our argument covers Coulomb interactions and does not need any Taylor expansion in time. It instead requires analyticity in space for the density and potential, which is compatible with extended nuclei.
\end{abstract}

\maketitle

Time-dependent density functional theory (TDDFT) \cite{RunGro-84,MarUllNogRubBurGro-06,Ullrichs-11,MarMaiNogGroRub-12} is the method of choice for the description of charge transfer in complex molecules~\cite{FolMauHamJayWahRagJonDimGaaSchLop-JPCA-23}, electronic response of large systems~\cite{JakLuBriLinSumGanBer-JCTC-25}, and ultrafast phenomena in attosecond physics~\cite{SatHubGioRub-NPJCM-25}. Unfortunately, its theoretical foundations are still under debate~\cite{MaiTodWooBur-10,FouLamLewSor-16,DFT-22,MAQUI_TDDFT-26} and a complete rigorous justification has only been given for finite discrete systems \cite{LiUll-08,FarTok-12,RugPenLee-15,MAQUI_finite-26}. 
In the continuum, both the Runge--Gross theorem~\cite{RunGro-84} about the uniqueness of the potential and the van Leeuwen expansion~\cite{vanLeeuwen-99} rely on the assumption that the potential is Taylor-expandable in time (real-analytic). However, this is not expected to hold with two-electron Coulomb interactions and would require unreasonable smoothness assumptions on initial states~\cite{MaiTodWooBur-10,YanMaiBur-12,YanBur-13,FouLamLewSor-16,LauLewOld-V-26_ppt}.
To circumvent these issues, fixed-point techniques bypassing Taylor-expandability were later proposed~\cite{MaiTodWooBur-10,RugLee-11,RugGiePenLee-12,RugPenLee-15,TarUll-21} and studied numerically \cite{RugPenLee-15,RugGiePenLee-12,NieRugLee-13}, but the theoretical validity of such techniques has remained an open question until now.

In this Letter, we provide the first entirely rigorous justification of TDDFT in the continuum, showing both existence and uniqueness of the potential, and consequently establish the rigorous foundation of Kohn--Sham theory \cite{KohSha-PR-65}. Our proof is valid for two-electron Coulomb interactions and both fermions and bosons. We work with minimal regularity assumptions in time, thus avoiding Taylor expansions in $t$ completely. 
Our densities and potentials are, however, required to be real-analytic in space (we only consider extended nuclei). 
We handle the singular  
Coulomb interaction using the real-analyticity of the wavefunction in one well-chosen direction.

We here only state the results, describe the physical implications, and provide the main ideas. The mathematical details can be found in the companion paper~\cite{LauLewOld-V-26_ppt}.
To show the physical consistency of our assumptions, we first consider the direct (potential-to-density) problem.

\medskip

\noindent{\bf Direct problem.} We work on the unit torus, that is, $\Lambda=[0,1]^3$ with periodic boundary conditions. Several of our arguments also hold on the whole space $\R^3$, but not all. The $N$-particle Hamiltonian in atomic units is 
\begin{equation*}
\cH_V:=-\sum_{j=1}^N\frac{\nabla^2_{\br_j}}2+\sum_{j=1}^N V(\br_j)+\sum_{1\leq j<k\leq N} V_{\rm ee}(\br_j-\br_k).
\end{equation*}
In practice $V_{\rm ee}$ is often the $\Z^3$-periodic Coulomb interaction, whose Fourier coefficients are $4\pi/k^2$ with $0\neq \bk\in2\pi\Z^3$. Here we want to keep $V_{\rm ee}$ arbitrary in order to cover all situations of interest, for instance $V_{\rm ee}=0$ in Kohn--Sham theory \cite{KohSha-PR-65}. We will only assume that 
\begin{equation}
    \int_\Lambda V_{\rm ee}(\br)^2\,\rd\br=\sum_\bk \bigl\vert \widehat{V_{\rm ee}}(\bk) \bigr\vert^2\quad\text{is finite,}
    \label{eq:Vee}
\end{equation}
which covers a much larger class of singular potentials than the periodic Coulomb repulsion.

The density $\rho_\Psi$ inherits the regularity of $\Psi$ but is usually smoother. For instance, ground state wavefunctions possess electron-electron cusps due to the singularity of the Coulomb repulsion, but their densities are nevertheless real-analytic away from the nuclei~\cite{FouHofOstHofOstSor-04,Jecko-10}. 
The intuition behind the smoothness is that the singularities of $\Psi$ are eliminated when we integrate over $N-1$ variables. We now present a very simple mechanism to formalize this intuition, which also helps in solving the inverse (density-to-potential) problem. 

We start with the observation that, for $V=0$, the Hamiltonian $\cH_0 = \cH_{V=0}$ is translation-invariant, hence commutes with the total momentum $\cP=-i\sum_{j=1}^N\nabla_{\br_j}$. This gives a particular role to the direction of the center of mass $\br_1+\cdots+\br_N$,
in which we can trust that $\Psi$ is smooth \cite{Hunziker-86}. For instance, the eigenstates of $\cH_0$ are all real-analytic in this direction because they are eigenfunctions of $\cP$. Similarly, the unitary group $e^{-it\cH_0}$ commutes with $\cP$ and will therefore propagate the real-analyticity of the initial condition to later times. 

The regularity of $\Psi$ in the direction $\br_1+\cdots+\br_N$ is automatically transferred to the density $\rho_\Psi$, because
\begin{multline}
\nabla_{\br_1}\rho_\Psi(\br_1)
=\nabla_{\br_1} N\!\idotsint_{\Lambda^{N-1}}\!|\Psi(\br_1,...,\br_N)|^2\rd\br_2\cdots\rd\br_N    \\
=N\!\idotsint_{\Lambda^{N-1}}\!\!\left(\sum_{j=1}^N\nabla_{\br_j}\right)\!|\Psi(\br_1,...,\br_N)|^2\rd\br_2\cdots\rd\br_N    
\label{eq:differentiate_density}
\end{multline}
and similarly for higher derivatives. Here we used that the inserted $\nabla_{\br_j}$'s yield a vanishing integral. The relation~\eqref{eq:differentiate_density} shows that any wavefunction $\Psi$ that is real-analytic in $\br_1+\cdots+\br_N$ automatically has a real-analytic density $\rho_\Psi$. The details are provided in the End Matter.
These observations suggest finding conditions on the external potential $V$ so that the real-analyticity of $\Psi$ in the direction $\br_1+\cdots+\br_N$ persists. The following states that it is sufficient to require $V(t,\br)$ to be real-analytic in $\br$ (extended nuclei are covered, but not point nuclei).

\begin{theorem}\label{thm:direct-V} Assume the interaction $V_{\rm ee}$ satisfies~\eqref{eq:Vee}. 

\smallskip

\noindent \emph{(Eigenstates).} Let $V(\br)$ be real-analytic. Then any eigenstate $\Psi$ of $\cH_V$
\vspace{-0.2cm}
\begin{subequations}
\label{eq:regular}
\begin{empheq}[left=\empheqlbrace]{align}
    &\text{is real-analytic in $\br_1+\cdots+\br_N$ and }
    \label{eq:regular.analytic}
    \\
    &\text{satisfies $\int_{\Lambda^N}|\nabla_{\br_j}^2\Psi|^2<\ii$ for $j=1,...,N$.}
    \label{eq:regular.H2}
\end{empheq}
\end{subequations}

\vspace{-0.2cm}
\noindent Its one-particle density $\rho_\Psi(\br)$ is real-analytic in $\br$. 

\smallskip

\noindent\emph{(Dynamics).} Let $V(t,\br)$ be continuous in $t\in[0,T]$ and real-analytic in $\br$. Let $\Psi_0$ satisfy~\eqref{eq:regular}. Then the solution $\Psi(t)$ to the time-dependent Schrödinger equation 
\begin{equation}
i\partial_t\Psi(t)=\cH_{V(t)}\Psi(t),\qquad \Psi(0)=\Psi_0
\label{eq:Schrodinger}
\end{equation}
is continuous in time and satisfies~\eqref{eq:regular} for all $t\in[0,T]$. The corresponding density $\rho_{\Psi}(t,\br)$ is twice-differentiable in $t$ and real-analytic in $\br$. 
\end{theorem}

For the inverse ($\rho\mapsto V$) problem studied below, it will be very important that Theorem~\ref{thm:direct-V} places $\rho_\Psi$ on the same footing of space regularity as $V$. Further, it will be crucial that the density is twice differentiable in time for potentials that are only continuous in time.

The real-analyticity of $\rho_\Psi$ provided in the theorem follows from that of $\Psi$ in $\br_1+\cdots+\br_N$ by~\eqref{eq:differentiate_density}. 
The property in~\eqref{eq:regular.H2}
gives some weak form of regularity in the other directions. It says that $\Psi$ not only has a finite kinetic energy, but also finite variance. Not much more can be expected for a singular interaction $V_{\rm ee}$.

Although we are mainly interested in the dynamics, we have mentioned eigenstates because $\Psi_0$ is often assumed to be a ground state, thereby justifying our assumptions at $t=0$. The analyticity-in-space of $\rho_{\Psi(t)}(\br)$ was conjectured in \cite{Zheng-07,Zheng-11} and is sometimes called the \emph{time-dependent holographic electron density theorem}. This plays an important role in TDDFT for open systems because it implies, by uniqueness of analytic continuation, that the knowledge of $\rho_\Psi$ in a subdomain at any time $t$ implies the knowledge of the whole density.

Next, we describe the main ideas of the proof of the dynamical part of the theorem (the missing mathematical technicalities can be found in~\cite{LauLewOld-V-26_ppt}). We recall that a function is real-analytic when its Taylor series has a positive radius of convergence at any point of space. This is equivalent to its Fourier coefficients decaying exponentially (the rate of decay is related to the radius of convergence in space), see the End Matter. Such a function being continuous in time requires this radius to be uniform in time. 

For an $N$-particle wavefunction $\Psi$, the real-analyticity in the direction of the center of mass thus means that the following series
\begin{equation}
    \pscal{\Psi,e^{2\sigma|\cP|}\Psi}=\sum_{\bk_j\in2\pi\Z^3}e^{2\sigma\left|\sum_{j=1}^N\bk_j\right|}|\widehat{\Psi}(\bk_1,...,\bk_N)|^2
    \label{eq:analyticity_Psi}
\end{equation}
is convergent for some $\sigma>0$ \footnote{To see this, note that the Fourier coefficient associated to the center of mass direction is $\bk_1+\ldots+\bk_N$, since $\cP$ is the generator of translations in this direction}. 
We assume this for $\Psi_0$ and then need to show this remains true for $\Psi(t)$ with a $\sigma$ uniform in time.

The property in~\eqref{eq:regular.H2}
requires an additional weight $\sum_{j=1}^N k_j^4$, but we neglect it here for brevity and refer to~\cite{LauLewOld-V-26_ppt} for the  details.
We differentiate in time the previous expectation and, after using that the commutator $[\cH_0,e^{2\sigma|\cP|}]=0$, since $\cH_0$ is translation-invariant, we find 
\begin{align}
    \frac{\rd}\dt \big\langle\Psi(t),e^{2\sigma|\cP|}\Psi(t)\big\rangle & =\big\langle\Psi(t),i\bigl[\cH_{V(t)},e^{2\sigma|\cP|}\bigr]\Psi(t)\big\rangle 
    \nn
    \\
    & = 2\Im \big\langle \chi(t),\cB(t) \chi(t)\big\rangle
    , 
    \label{eq:commutator}
\end{align}
with $\chi := e^{\sigma|\cP|}\Psi$ and $\cB:=e^{\sigma|\cP|}\sum_{j=1}^NV(\br_j)e^{-\sigma|\cP|}$. We then bound 
\begin{multline*}
    \left|\pscal{\chi(t),\cB(t) \chi(t)}\right|\\
    \leq N\sum_{\bk_1,\bk_1',\bk_2\cdots}e^{\sigma|\bk_1+\cdots+\bk_N|}|\widehat{V}(t,\bk_1-\bk_1')|e^{-\sigma|\bk_1'+\cdots+\bk_N|}\\
    \times|\widehat{\chi}(t,\bk_1,\bk_2,...,\bk_N)|\,|\widehat{\chi}(t,\bk'_1,\bk_2,...,\bk_N)|.
\end{multline*}
Using the triangle inequality $|\bk_1+\cdots+\bk_N|\leq |\bk_1-\bk_1'|+|\bk_1'+\cdots+\bk_N|$ and calling $\sigma_V$ the (uniform in time) exponential decay rate for $\widehat{V}(t,\bk)$, we find
\begin{multline*}
e^{\sigma|\bk_1+\cdots+\bk_N|}|\widehat{V}(t,\bk_1-\bk_1')|e^{-\sigma|\bk_1'+\cdots+\bk_N|}\\
\leq e^{\sigma|\bk_1-\bk_1'|}|\widehat{V}(t,\bk_1-\bk_1')|\leq  C e^{-(\sigma_V-\sigma)|\bk_1-\bk_1'|}.
\end{multline*}
We choose $\sigma<\sigma_V$ and use the bound $2ab \leq a^2 + b^2$ with $a = |\widehat{\chi}(t,\bk_1,\bk_2,...,\bk_N)|$ and $b= |\widehat{\chi}(t,\bk_1',\bk_2,...,\bk_N)|$. Computing the sums we obtain 
$$\left|\pscal{\chi,\cB(t)\chi}\right|\leq CN\pscal{\chi,\chi}\sum_\bk e^{-(\sigma_V-\sigma)|\bk|}.$$
This proves that the term on the right of~\eqref{eq:commutator} is less than $C'\pscal{\Psi(t),e^{2\sigma|\cP|}\Psi(t)}$ for some constant $C'$ and after integration in time that
\begin{equation*}
    \big\langle\Psi(t),e^{2\sigma|\cP|}\Psi(t)\big\rangle\leq e^{C't}\big\langle\Psi_0,e^{2\sigma|\cP|}\Psi_0\big\rangle,
\end{equation*}
which is the claimed preservation of the real-analyticity in the center of mass in a real-analytic external potential $V$. 
The relation~\eqref{eq:differentiate_density} then implies the real-analyticity of $\rho_\Psi$, see also the End Matter. 

The proof that $\rho_\Psi$ admits two continuous derivatives in $t$ is more complicated~\cite{LauLewOld-V-26_ppt}. The first derivative satisfies the \emph{continuity equation}
\begin{equation*}%
    \partial_t \rho_{\Psi(t)} = - \nabla_{\mathbf{r}} \cdot j_{\Psi(t)}, 
\end{equation*}
where $j_\Psi(\br) = N \Im \int_{\Lambda^{N-1}} \overline{\Psi(\br,\bR)}\nabla_{\br}\Psi(\br,\bR) \,\rd\bR$ with $\bR=(\br_2,...,\br_N)$ is the current of $\Psi$. 
The second derivative satisfies
the \emph{force-balance equation} also known as the (divergence of the) \emph{local force equation}
\cite{vanLeeuwen-99,TchPenTheRugRub-19,TanPenLaeCsiRugRub-24,TanPenRugRub-25}
\begin{equation}\label{eq:force-balance}
    \partial_t^2 \rho_{\Psi(t)} 
		= - \frac{\nabla_\br^4\rho_{\Psi(t)}}{4} +\nabla_\br\cdot(\rho_{\Psi(t)}\nabla_\br V)
        + Q_{\Psi(t)},
\end{equation}
where
\begin{multline}
    Q_\Psi(\br) 
    = N\sum_{\alpha,\beta=1}^3 \partial_\br^\alpha\partial_\br^\beta\int_{\Lambda^{N-1}}\overline{\partial_\br^\alpha \Psi(\br,\bR)} \partial_\br^\beta\Psi(\br,\bR)\,\rd\bR\\
	    -2N\Re\nabla_\br  \cdot \int_{\Lambda^{N-1}} \overline{\Psi(\br,\bR)}\nabla_{\br}\Psi(\br,\bR)\sum_{j=2}^N V_{\rm ee}(\br-\br_j)\,\rd\bR\\
        + N\nabla^2_\br  \int_{\Lambda^{N-1}}  |\Psi(\br,\bR)|^2   \sum_{j=2}^N V_{\rm ee}(\br-\br_j)\,\rd\bR
        .
    \label{eq:Q4}
\end{multline}
In 
\eqref{eq:Q4}, $\partial^\alpha_\br$ is the derivative in the direction $r^\alpha\in\R$ for $\br=(r^1,r^2,r^3)$. 
\Cref{eq:force-balance} gives an explicit link between $\Psi$ and $V$, and thus plays a central role in TDDFT~\cite{vanLeeuwen-99}.
We have written it with as many derivatives outside of the integrals over the $N-1$ remaining particles as possible. In particular, no individual $\Psi$ in $Q_\Psi$ appears with four derivatives. 
This will be important later.

It follows from the same reasoning as in~\eqref{eq:differentiate_density} applied to the three integrals in~\eqref{eq:Q4} that $Q_\Psi$ is also a real-analytic function. Indeed, any additional $\nabla_{\br}$ can be transformed into $\nabla_\br+\sum_{j=2}^N\nabla_{\br_j}$ and  commuted with the interaction $V_{\rm ee}$ by translation-invariance. In this manner we again end up differentiating terms involving $\Psi$ only and we can use its real-analyticity in the direction $\br_1+\cdots+\br_N$. This implies the twice differentiability of $\rho_\Psi$ in $t$.

\medskip

\noindent{\bf Inverse problem.}
We next turn to the inverse problem of finding a potential $V(t,\br)$ such that the corresponding Schrödinger equation reproduces a given density $\rho(t,\br)$, which we solve for \emph{any} interaction $V_{\rm ee}$ satisfying \eqref{eq:Vee}.

\begin{theorem}[Inverse problem in TDDFT]\label{thm:inverse-V} 
Assume the interaction $V_{\rm ee}$ satisfies~\eqref{eq:Vee}. Let the initial state $\Psi_0$ satisfy~\eqref{eq:regular}. Consider a density trajectory $\rho(t,\br)$ that is twice differentiable in $t$, real-analytic in $\br$, satisfies 
$\rho(t=0)=\rho_{\Psi_0}$, 
$\partial_t\rho(t=0) = -\nabla_\br \cdot j_{\Psi_0}$, 
$\int_\Lambda \rho(t,\br) \, \rd \br = N$, and 
$\rho(t,\br)>0$ for all $0\leq t\leq T$ and $\br\in\Lambda$.

Then, there exists some time $0<T'\leq T$ and an external potential $V(t,\br)$ that is continuous in $t$ and real-analytic in $\br$, unique up to addition of a constant $C(t)$, such that the solution $\Psi(t)$ to  the Schrödinger equation~\eqref{eq:Schrodinger} has $\rho_{\Psi(t)}(\br)=\rho(t,\br)$ for $t\leq T'$.
\end{theorem}

Since $\rho(0,\br)>0$, the strict positivity of $\rho(t,\br)$ is automatic by continuity, if we choose $T$ small enough. The regularity assumptions on $\rho$ and $V$ exactly match those in \cref{thm:direct-V}. We have thus established a one-to-one correspondence between physically relevant classes of densities and potentials, for any given $\Psi_0$ satisfying~\eqref{eq:regular} and $\rho_{\Psi_0}>0$. This is called the (unique) $v$-representability problem in TDDFT and it is for the first time solved here. The assumptions at time $t=0$ are necessary to have compatibility between $\rho$ and $\Psi_0$. The additional condition that $\rho > 0$ is technical and usually expected to hold if $\Psi_0$ is a ground state.

The class of $v$-representable densities determined by \cref{thm:inverse-V} does \emph{not} depend on the interaction $V_{\mathrm{ee}}$. In particular, this implies that one can uniquely represent the true $N$-particle Coulomb density by non-interacting fictitious electrons as is done in Kohn--Sham theory ($V_{\rm ee}=0$), or by a system with a damped interaction for the adiabatic connection \cite[Chs.~13--14]{Ullrichs-11}. 
Even for analytic interactions, where the wavefunction $\Psi$ is analytic in all its variables, the rigorous construction of the inverse potential $V$ is first achieved here.

The inverse potential $V$ depends continuously on the initial state $\Psi_0$ and the density $\rho$. Concretely, if $\tilde\rho$ is another density satisfying the assumptions with $\rho-\tilde\rho$ small enough, the potential $\tilde V$ obtained from the density $\tilde\rho$
satisfies $\|\tilde V-V\|\leq C \|\rho-\tilde\rho\|+C\|\partial_t^2\rho-\partial_t^2\tilde\rho\|$ 
with the norms considered in~\cite{LauLewOld-V-26_ppt} and in the End Matter (for appropriate $\sigma$'s). This can be viewed as a `quantitative Runge--Gross theorem' analogously to the recent work in the static case (i.e.~Hohenberg--Kohn theorem \cite{HohKoh-64}) primarily in one dimension~\cite{SutSarPenRugLeeGie-24,Corso-2025,CorLae-25,Corso-26_ppt}.

\cref{thm:inverse-V} allows us to rigorously define the universal functional potential (with or without $V_{\mathrm{ee}}$) depending on the initial state and the density trajectory, and hence the Hartree-exchange-correlation (Hxc) functional following the standard construction \cite{MarMaiNogGroRub-12,Ullrichs-11,MarUllNogRubBurGro-06,LauLewOld-V-26_ppt}. This gives the first rigorous foundation of time-dependent Kohn--Sham theory, where the true $N$-particle Coulomb density is reproduced from self-consistent Kohn--Sham equations.

The rigorous proof of Theorem~\ref{thm:inverse-V} is involved and can be found in~\cite{LauLewOld-V-26_ppt}. We present here the main ideas. 
For a given trajectory $\rho(t,\br)$ and a given wavefunction $\Psi$, we define a potential $V[\Psi,\rho](t,\br)$ through Eq.~\eqref{eq:force-balance} by
\begin{equation}\label{eq:force-balance_nonlinear}
    \nabla_\br\cdot\big(\rho(t)\nabla_\br V[\Psi,\rho]\big)=\partial_t^2 \rho +\frac{\nabla^4_\br\rho}4- Q_{\Psi}.
\end{equation}
This is the reason we need that $\rho$ is twice differentiable in $t$. Further, it is very important that we have replaced all the $\rho_\Psi$'s by the prescribed density $\rho$, in particular the term $\nabla^4_\br\rho_{\Psi}$ that would otherwise involve four derivatives of $\Psi$. This replacement is new and is at the heart of the proof. The linear equation~\eqref{eq:force-balance_nonlinear} has a unique well-defined solution $V=V[\Psi,\rho]$ such that $\int_\Lambda V(t,\br)\,\rd\br=0$, whenever $\rho$ is strictly positive on the torus $\Lambda$. This is because the operator $K_\rho(V)= -\nabla_\br\cdot(\rho\nabla_\br V)$ is elliptic and written in divergence form. Its associated quadratic form $\int_\Lambda \rho|\nabla_\br V|^2 \, \rd \br$ is  strictly positive on mean-zero functions; note that the right-hand-side of \eqref{eq:force-balance_nonlinear} integrates to 0. Our assumption that $\rho(t,\br)$ is strictly positive is to ensure solvability of this linear problem.

Equation~\eqref{eq:force-balance_nonlinear} allows us to reformulate the inverse problem as a nonlinear Schrödinger equation in the wavefunction $\Psi$, of the form
\begin{equation}
i\partial_t\Psi(t)=\cH_{V[\Psi(t),\rho]}\Psi(t),\qquad \Psi(0)=\Psi_0.
\label{eq:nonlinear_Psi}
\end{equation}
If we can solve this highly nonlinear equation, then the sought-after potential will simply be $V=V[\Psi,\rho]$. Indeed, by \eqref{eq:force-balance}, \eqref{eq:force-balance_nonlinear}, and the compatibility conditions, the difference $h=\rho_\Psi - \rho$ between the resulting density $\rho_\Psi$ and the prescribed one $\rho$ satisfies 
\begin{equation*}
\partial_t^2 h  =
-\frac{\nabla^4_\br h}4 + \nabla_\br\cdot\big(h\nabla_\br V\big), 
\qquad \partial_th(0)=h(0)=0. 
\end{equation*}
The unique solution to this wave-like equation is $h=0$, whence $\rho_\Psi = \rho$ \cite{LauLewOld-V-26_ppt}.

\Cref{eq:nonlinear_Psi} is in the spirit of~\cite{MaiTodWooBur-10} but with a different functional $V[\Psi,\rho]$ due to the $\nabla^4_\br\rho$ term. Proving the well-posedness of~\eqref{eq:nonlinear_Psi} is a hard mathematical question due to a phenomenon of \emph{loss of derivatives}, which prevents the use of standard fixed-point methods \footnote{The \emph{loss of derivatives} refers to the fact that \eqref{eq:nonlinear_Psi} involves nonlinear terms with spatial derivatives of $\Psi$. 
In a time-discretization, this means that $\Psi(t+\delta t)$ admits fewer spatial derivatives than $\Psi(t)$. With $\delta t\to 0$ we see that $\Psi(t')$ admits infinitely many fewer derivatives than $\Psi(t)$ for any $t'>t$. (The linear part $\cH_0$ can be treated using the unitary group $e^{-it\cH_0}$ and does not lead to any loss.)}

In principle, the function $Q_\Psi$ involves four space derivatives, and this is where our specific rewriting~\eqref{eq:force-balance}--\eqref{eq:Q4} is essential. When we invert the elliptic operator $K_\rho$ on the left of~\eqref{eq:force-balance_nonlinear} to find $V$, we gain two derivatives and thus compensate the derivatives outside the integrals in \eqref{eq:Q4}. With this regularizing effect, we see that $V[\Psi,\rho]$ morally has a loss of only one derivative. 
This is good news because, although there exists no tool to deal with general nonlinear Schrödinger equations losing multiple derivatives, the case of one lost derivative is more under control. 
(Had we not replaced the term $\nabla_{\br}^4\rho_\Psi$ by $\nabla_{\br}^4\rho$ in \eqref{eq:force-balance_nonlinear}, we would morally have a loss of two derivatives instead. This is why the particular way we write 
\eqref{eq:force-balance}--\eqref{eq:Q4} is important.)

A classical tool to find solutions to equations losing one derivative is the \emph{Cauchy--Kowalevski theorem}~\cite{Nirenberg-72,Nishida-77}, which works in the real-analytic setting. 
The main difficulty in our case is that $\Psi$ is analytic only in the direction $\br_1+\cdots+\br_N$, which complicates the analysis dramatically (analytic interactions $V_{\rm ee}$ are much easier).
The adaptation of the Cauchy--Kowalevski theorem to the present setting is explained in detail in \cite{LauLewOld-V-26_ppt}. The  Cauchy--Kowalevski approach was used for classical mean-field TDDFT (Vlasov--Poisson equation \cite{ChaFin-05,Manfredi-20}) independently and simultaneously in \cite{Bouedec-26_ppt}.

\medskip

\noindent{\bf Iteration schemes.} The Cauchy--Kowalevski theorem also yields an iteration scheme converging to the desired $V$. 
The idea is to solve iteratively
\begin{equation}
i\partial_t\Psi_{n+1}(t)=\cH_{0}\Psi_{n+1}(t)+\sum_{j=1}^NV[\Psi_n(t),\rho](t,\br_j)\Psi_n(t),
\label{eq:iteration_Psi_source}
\end{equation}
where the entire nonlinearity is seen as a source term. (The iteration starts at $\Psi_0(t) = e^{-it\cH_0}\Psi_0$ and for each $n$, the potential is found from \eqref{eq:force-balance_nonlinear} with $\Psi=\Psi_n$.) The standard approach~\cite{Nirenberg-72,Nishida-77} to show convergence of $\Psi_n$ to the solution of the problem is to slightly decrease the maximal existence time $T_n$ and the exponential decay rate $\sigma_n$ along the iteration, in a way that $T_n\to T_\infty>0$, $\sigma_n\to\sigma_\infty>0$ (cf.~the End Matter). In other words, the loss of a derivative is compensated by sacrificing a bit of existence time and analyticity radius for the Taylor series.
The associated potential $V[\Psi_n(t),\rho]$ converges to the unique inverse potential of Theorem~\ref{thm:inverse-V} that satisfies $\int_\Lambda V(t,\br)\,\rd\br=0$ for all $0\leq t\leq T_\infty$~\cite{LauLewOld-V-26_ppt}.

Following~\cite{RugLee-11,RugGiePenLee-12,NieRugLee-13,RugPenLee-15,TarUll-21} we can also consider the sequence
\begin{equation}
i\partial_t\tilde\Psi_{n+1}(t)=\cH_{V[\tilde\Psi_n(t),\rho]}\tilde\Psi_{n+1}(t).
\label{eq:iteration_Psi}
\end{equation}
This is more physical because the iterates $\Psi_n$ of~\eqref{eq:iteration_Psi_source} will not necessarily be normalized before reaching convergence due to the source term, whereas $\tilde\Psi_n$ are. This is not exactly the scheme proposed in~\cite{RugLee-11,RugGiePenLee-12,NieRugLee-13,RugPenLee-15,TarUll-21} because our functional $V[\Psi,\rho]$ defined in~\eqref{eq:force-balance_nonlinear} is different, due to the replacement of $\nabla_\br^4\rho_\Psi$ by $\nabla_\br^4\rho$. 
As mentioned above, this is crucial to guarantee convergence. We show in the End Matter that $\tilde\Psi_n-\Psi_n\to0$, ensuring convergence of~\eqref{eq:iteration_Psi} as well. We cannot prove convergence of the original scheme proposed in~\cite{RugLee-11,RugGiePenLee-12,NieRugLee-13,RugPenLee-15,TarUll-21}, since this corresponds to a nonlinear equation with a loss of two derivatives. 

\medskip

\noindent{\bf Conclusion.} 
We have provided the first fully rigorous solution to the $v$-representability problem in TDDFT. We work with potentials and densities that are real-analytic in space but have minimal regularity in time, which are physically relevant for Coulomb systems. The wavefunction is analytic in the direction of the center of mass. 
The proof relies on rewriting the problem as a nonlinear equation in $\Psi$ that essentially loses only one derivative. The sought-after potential can be obtained by an iteration scheme. 
All the missing mathematical details are described in the companion paper~\cite{LauLewOld-V-26_ppt}.

In future work, we will compare the two schemes~\eqref{eq:iteration_Psi_source} and~\eqref{eq:iteration_Psi} on practical examples and we will generalize our findings to the case of the whole space, where the invertibility of the elliptic operator $K_\rho$ is more involved. 

\medskip

\noindent
\textbf{Acknowledgment.} This work has benefited from French State support managed by ANR under the France 2030 program through the MaQuI CNRS Risky and High-Impact Research program (RI)$^2$ (grant agreement ANR-24-RRII-0001). We thank \'Eric Cancès, Théo Duez, Jari van Gog and Julien Toulouse for stimulating discussions within this project.

\bibliography{biblio_maqui}

\appendix
\section*{End Matter}

\noindent{\bf Details for the analyticity of the density.}
Using~\eqref{eq:differentiate_density}, it is possible to show that any $\Psi$ satisfying~\eqref{eq:analyticity_Psi} has a real-analytic density. We provide this proof in Fourier space, in the same spirit as the other arguments in the Letter. Using $|\bk|\leq |\bk+\bK|+|\bK|$ with $\bK=\bk_1+\cdots +\bk_N$ and $\sigma'<\sigma$ we can bound
\begin{align*}
&\frac{1}N\sum_{\bk}e^{\sigma'|\bk|}|\widehat{\rho_\Psi}(\bk)|\\
&\quad \leq \sum_{\bk,\bk_j}e^{\sigma'|\bK|}\bigl\vert\widehat{\Psi}(\bk_1,...,\bk_N)\bigr\vert e^{\sigma'|\bk+\bK|} \bigl\vert\widehat{\Psi}(\bk+\bk_1,...,\bk_N)\bigr\vert\\
&\quad \leq \sum_{\bk,\bk_j}e^{2\sigma|\bK|+2(\sigma'-\sigma)|\bk+\bK|}\bigl\vert\widehat{\Psi}(\bk_1,...,\bk_N)\bigr\vert^2\\
&\quad =\pscal{\Psi,e^{2\sigma|\cP|}\Psi}\sum_{\bk}e^{-2(\sigma-\sigma')|\bk|}.
\end{align*}
In the second line we used $2ab\leq a^2+b^2$ with $a=e^{\sigma|\bK|-(\sigma-\sigma')|\bk+\bK|}|\widehat{\Psi}(\bk_1,...,\bk_N)|$. The analyticity of $\rho_\Psi$ hence follows from that  of $\Psi$ in the direction of the center of mass. To see that the exponential decay in Fourier implies the convergence of the Taylor series, recall
\begin{equation*}
\partial^{\alpha_1}_{r^1}\partial^{\alpha_2}_{r^2}\partial^{\alpha_3}_{r^3}\rho_\Psi(\br) 
=\sum_{\bk}(ik^1)^{\alpha_1}(ik^2)^{\alpha_2}(ik^3)^{\alpha_3}\widehat{\rho_\Psi}(\bk)e^{i\bk\cdot\br}.
\end{equation*}
Using $x^n\leq \frac{n!}{R^n}e^{Rx}$ with $x=|k^1|,|k^2|,|k^3|$ as well as  $|k^1|+|k^2|+|k^3|\leq \sqrt3|\bk|$ in the exponential, we obtain
$$
\left|\partial^{\alpha_1}_{r^1}\partial^{\alpha_2}_{r^2}\partial^{\alpha_3}_{r^3}\rho_\Psi(\br)\right|
\leq  \frac{(\alpha_1!)(\alpha_2!)(\alpha_3!)}{R^{\alpha_1+\alpha_2+\alpha_3}}\sum_{\bk}e^{R\sqrt3|\bk|}|\widehat{\rho_\Psi}(\bk)|.
$$
This proves the convergence of the Taylor series with radius $R<\sigma'/\sqrt3$ about each point of the torus $\Lambda$.

\smallskip

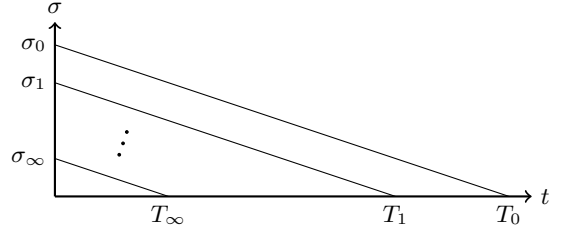
\begin{figure}
\centering
\begin{tikzpicture}
    \draw[->, thick] (0,0) -- (6.3,0) node[anchor=west] {$t$};
    \draw[->, thick] (0,0) -- (0,2.3) node[anchor=south] {$\sigma$};
    \draw (0,2) -- (6,0);
    \node[anchor=east] at (0,2) {$\sigma_0$};
    \node[anchor=north] at (6,0) {$T_0$};
    \draw (0,1.5) -- (4.5,0);
    \node[anchor=east] at (0,1.5) {$\sigma_1$};
    \node[anchor=north] at (4.5,0) {$T_1$};
    \foreach \x in {-1,0,1} 
        \filldraw (0.9+0.05*\x,0.7+0.15*\x) circle (0.5pt);
    \draw (0,0.5) -- (1.5,0);
    \node[anchor=east] at (0,0.5) {$\sigma_\infty$};
    \node[anchor=north] at (1.5,0) {$T_\infty$};
\end{tikzpicture}
\caption{The regularity parameter decreases linearly in time to absorb the loss of derivatives. Along the iteration scheme, both the initial regularity $\sigma_n$ and final time $T_n$ are decreased, but the slope is kept constant. The limit initial regularity $\sigma_\infty = \lim\sigma_n$ and final time $T_\infty = \lim T_n$ are strictly positive.\label{fig:triangle}}
\end{figure}

\noindent{\bf Convergence of the iteration scheme~\eqref{eq:iteration_Psi}.} In~\cite{LauLewOld-V-26_ppt} we work with the norm
$$\norm{\Psi}_\sigma=\big\|(A+\cH_0)\pscal{\cP}e^{\sigma|\cP|}\Psi\big\|_{L^2}$$
where $\pscal{\cP}=\sqrt{1+\cP^2}$ (Japanese bracket) and $\cP=-i\sum_{j=1}^N\nabla_{\br_j}$ is the total momentum. We choose $A$ so that $\cH_0+A\geq 1$. We also use the space-time norm
$$\norm{\Psi}_{T,\sigma}=\sup_{\substack{0\leq t\leq T\\ 0\leq \sigma'\leq \sigma(1-t/T)}}\norm{\Psi(t)}_{\sigma'},$$
which is defined on a ``triangle'' in the $(t,\sigma)$ plane allowing for a linear decrease of analytic regularity over time, see \cref{fig:triangle}.

We proved in~\cite{LauLewOld-V-26_ppt} that the sequence~\eqref{eq:iteration_Psi_source} satisfies
\begin{equation}
    \norm{\Psi_{n+1}-\Psi_n}_{T'_0,\sigma_0'}\leq \frac{C}{2^n},  \qquad  \norm{\Psi_n}_{T'_0,\sigma'_0}\leq C
    \label{eq:CV_Psi_n}
\end{equation}
for some constant $C$, provided $T_0'$ is small enough. The regularity parameter $\sigma'_0$ is also chosen smaller than the one of the prescribed density $\rho(t,\br)$.

Next, integrating the two Schrödinger equations~\eqref{eq:iteration_Psi_source} and~\eqref{eq:iteration_Psi} using that they start at the same $\Psi_0$, we find
\begin{multline}
\Psi_{n+1}(t)-\tilde \Psi_{n+1}(t)\\
=-i\int_0^te^{-i(t-s)\cH_0}\bigl(\cV_n\Psi_n-\tilde\cV_n\tilde\Psi_{n+1}\bigr)(s)\,\rd s
\label{eq:Duhamel}
\end{multline}
with $\cV_n:=\sum_{j=1}^NV[\Psi_n,\rho](t,\br_j)$ and similarly for $\tilde\cV_n$ with $\tilde\Psi_n$.
To account for the loss of derivatives included in $\tilde\cV_n$, we follow~\cite{Nirenberg-72,Nishida-77} and introduce a new space-time norm
\begin{equation}
\nnorm{\Psi}_{T,\sigma}=\sup_{\substack{0\leq t\leq T\\ 0\leq \sigma'\leq \sigma(1-t/T)}}\norm{\Psi(t)}_{\sigma'}\big(\sigma(1-t/T)-\sigma'\big)
\label{eq:nnorm}
\end{equation}
with blow-up allowed on the right side of the triangle.

We define the sequences $\sigma_{n+1}=(1-\frac{1}{2(n+1)^2})\sigma_n$ and $T_{n+1}=(1-\frac{1}{2(n+1)^2})T_n$ with (fixed) $\sigma_0<\sigma_0'$ and (small) $T_0<T_0'$ to be determined later. Note that $\sigma_n$ and $T_n$ have a positive limit because
$\prod_{n\geq1}\left(1-\frac{1}{2n^2}\right)\geq \exp({-\sum_{n\geq1}\frac1{n^2}}) > 0$. Note also that $\sigma_n/T_n=\sigma_0/T_0$ for all $n$, hence we are moving the right side of the triangle parallel to the original side, see Figure~\ref{fig:triangle}. We write $\sigma_n(t)=\sigma_n(1-t/T_n)=\sigma_n-\sigma_0t/T_0$ and denote for simplicity $\|\cdot\|_n:=\|\cdot\|_{T_n,\sigma_n}$ and similarly for~\eqref{eq:nnorm}. We then prove by induction that
\begin{equation}\label{eq:induction}
\big\|\Psi_n-\tilde\Psi_n\big\|'_{n}\leq \frac{M}{2^n},\qquad \|\tilde\Psi_n\|_{n}\leq M
\end{equation}
when $T_0$ is small enough and for some well chosen $M > 1$. We thus assume the validity of this bound and prove it for $n+1$. Using~\eqref{eq:Duhamel} we first estimate
(suppressing the time in the notation)
\begin{multline*}
\|\tilde\Psi_{n+1}\|_{\sigma'}\leq\norm{\Psi_{n+1}}_{\sigma'}+T_0\norm{\cV_n}_{\sigma'}\norm{\Psi_n}_{\sigma'}\\ 
+T_0 \|\cV_{n} \|_{\sigma'} \|\tilde\Psi_{n+1}\|_{\sigma'}
+T_0 \|\tilde\cV_{n}-\cV_{n}\|_{\sigma'}\|\tilde\Psi_{n+1}\|_{\sigma'}.
\end{multline*}
(Here the norm on $\cV$'s is the operator norm arising from the norm $\|\cdot\|_{\sigma'}$.)
Terms involving $\cV_n$ and $\Psi_n$ are bounded~\cite{LauLewOld-V-26_ppt}. Furthermore, estimates from~\cite{LauLewOld-V-26_ppt} imply 
\begin{multline}\label{eq:bdd.VtildeminusV}
\big\| \tilde\cV_{n} - \cV_n \big\|_{\sigma'}
\leq C
    \big(1+\|\tilde\Psi_n\|_{\sigma'}\big)
    \big\|\pscal{\cP}(\Psi_n-\tilde\Psi_n)\big\|_{\sigma'}
\\ 
  + C \big(1 + \Vert\pscal{\cP}\tilde\Psi_n\Vert_{\sigma'}\big)
 \Vert\Psi_n - \tilde\Psi_n\Vert_{\sigma'}
 .
\end{multline} 
(Everywhere $C$ designates a generic constant that can change from line to line and does not depend on the sequence $\tilde\Psi_n$; we do, however, absorb $\sigma_0$ in the constant.)
We bound the first term. The second follows analogously.
Using $\pscal{x} e^{\sigma' |x|}\leq C \frac{e^{\sigma |x|}}{\sigma-\sigma'}$ for $\sigma'<\sigma$ and the induction hypothesis, we can bound
\begin{multline}
\big\|\pscal{\cP}(\Psi_n-\tilde\Psi_n)(t)\big\|_{\sigma_{n+1}(t)}
\leq \frac{\big\|(\Psi_n-\tilde\Psi_n)(t)\big\|_{\sigma''}}{\sigma''-\sigma_{n+1}(t)}
\\
\leq \frac{\|\Psi_n-\tilde\Psi_n\|'_n}{\big(\sigma''-\sigma_{n+1}(t)\big)\big(\sigma_n(t)-\sigma''\big)}.\label{eq:loss}
\end{multline}
We choose $\sigma''=(\sigma_n(t)+\sigma_{n+1}(t))/2$ and obtain from \eqref{eq:induction}
\begin{align*}
\big\|\pscal{\cP}(\Psi_n-\tilde\Psi_n)(t)\big\|_{\sigma_{n+1}(t)}
&\leq \frac{16M(n+1)^4}{\sigma_n^2 2^{n}}\leq CM. 
\end{align*}
Thus, by the induction hypothesis \eqref{eq:induction} and taking the supremum over $\sigma'$, we have
$\|\tilde\cV_n-\cV_n\|_{n+1} 
    \leq CM^2$
(norm being operator norm) and thus 
\begin{equation*}
\|\tilde\Psi_{n+1}\|_{n+1}
\leq C(1+T_0)
+CT_0 M^2\|\tilde\Psi_{n+1}\|_{n+1}. 
\end{equation*}
We thus take $M>1$ sufficiently large and $T_0$ small so that 
$\|\tilde\Psi_{n+1}\|_{n+1}\leq M$ as we wanted.

Recalling next \eqref{eq:Duhamel}, we write
\begin{multline*}
\cV_n\Psi_n-\tilde\cV_{n}\tilde\Psi_{n+1}
=\cV_n(\Psi_n-\Psi_{n+1})
+(\cV_n-\tilde\cV_{n})\tilde\Psi_{n+1}\\
+\cV_n(\Psi_{n+1}-\tilde\Psi_{n+1})
.
\end{multline*}
 Inserting in \eqref{eq:Duhamel} and using \eqref{eq:CV_Psi_n} we obtain, similarly as before, for $\sigma'<\sigma_{n+1}(t)$,
\begin{multline}
\big\|\Psi_{n+1}(t)-\tilde\Psi_{n+1}(t)\big\|_{\sigma'}
 \\ \leq
C\frac{T_0}{2^n}
+C M \int_0^t\big\|(\cV_n-\tilde\cV_{n})(s)\big\|_{\sigma'} \, \rd s\\
+C\int_0^t\big\|(\Psi_{n+1}-\tilde\Psi_{n+1})(s)\big\|_{\sigma'} \, \rd s 
.\label{eq:diff.Psi.level.n+1}
\end{multline}
To bound the second term we use \eqref{eq:induction} and \eqref{eq:bdd.VtildeminusV} and bound the integral $\int_0^t \norm{\pscal{\cP}(\Psi_n - \tilde\Psi_n)(s)}_{\sigma'} \rd s$. (The other term from \eqref{eq:bdd.VtildeminusV} is bounded similarly.) 
Arguing as in~\eqref{eq:loss} we have with $\sigma''(s)=(\sigma'+\sigma_n(s))/2$
\begin{align*}
&\int_0^t\big\|\pscal{\cP}(\Psi_n-\tilde\Psi_n)(s)\big\|_{\sigma'}\rd s\\
&\quad\leq\big\|\Psi_n-\tilde\Psi_n\big\|_{n}'\int_0^t\frac{\rd s}{(\sigma''(s)-\sigma')(\sigma_n(s)-\sigma''(s))}\\
&\quad = \frac{4T_0 \|\Psi_n - \tilde\Psi_n\|_n'}{\sigma_0}
\left(\frac{1}{\sigma_n(t) - \sigma'} - \frac{1}{\sigma_n-\sigma'}\right)
\\ & \quad
\leq\frac{4MT_0}{\sigma_02^n(\sigma_{n+1}(t)-\sigma')}
\end{align*}
by the induction hypothesis \eqref{eq:induction}. 
Similarly, the third term in \eqref{eq:diff.Psi.level.n+1} is bounded by 
\begin{align*}
    & \int_0^t\big\|(\Psi_{n+1}-\tilde\Psi_{n+1})(s)\big\|_{\sigma'} \, \rd s 
    \\ & \quad \leq 
    \frac{T_0\big\|\Psi_{n+1}-\tilde\Psi_{n+1}\big\|_{n+1}'}{\sigma_0}
    \log \frac{\sigma_{n+1} - \sigma'}{\sigma_{n+1}(t) -\sigma'}
    .
\end{align*}
Inserting these in \eqref{eq:diff.Psi.level.n+1}, multiplying by $\sigma_{n+1}(t)-\sigma'$, and taking the supremum as is needed for the norm~\eqref{eq:nnorm}, we find
\begin{equation*}
\big\|\Psi_{n+1}-\tilde\Psi_{n+1}\big\|_{n+1}'
\leq
C\frac{M^3T_0}{2^n}
+C T_0 \big\|\Psi_{n+1}-\tilde\Psi_{n+1}\big\|_{n+1}'
.
\end{equation*}
For $T_0$ small enough we arrive at the claimed bound $\|\Psi_{n+1}-\tilde\Psi_{n+1}\|_{n+1}'\leq M / 2^{n+1}$.
This proves by induction the bounds \eqref{eq:induction}.

To finally conclude the convergence of the sequence $\tilde \Psi_n$, let $\tilde T = T_\infty/2$, $\tilde\sigma = \sigma_\infty/2$, and let $\Psi_\infty$ denote the limit of the $\Psi_n$'s (in the $\norm{\cdot}_{T_0',\sigma_0'}$-norm). Then,
\begin{align*}
    \norm{\tilde \Psi_n - \Psi_\infty}_{\tilde T, \tilde \sigma} 
    & \leq \norm{\tilde \Psi_n - \Psi_n}_{\tilde T, \tilde \sigma}  + 
    \norm{\Psi_n - \Psi_\infty}_{\tilde T, \tilde \sigma} 
    \\ & \leq \frac{1}{\tilde\sigma} \norm{\tilde \Psi_n - \Psi_n}_{n}' + 
    \norm{\Psi_n - \Psi_\infty}_{T_0', \sigma_0'}.
\end{align*}
By \eqref{eq:CV_Psi_n} and \eqref{eq:induction} this proves that $\tilde \Psi_n$ converges to $\Psi_\infty$ in the $\norm{\cdot}_{\tilde T, \tilde \sigma}$-norm.

\end{document}